\documentclass[aps,prd,groupedaddress]{revtex4}
\usepackage{amsmath}
\usepackage{amssymb}
\usepackage{bbm}
\usepackage{epsfig}
\allowdisplaybreaks[4]
\begin{document}

\title{Contributions from dimension six strong flavor changing
operators to $t\bar{t}$, $t$ plus gauge boson, and $t$ plus Higgs boson production at the LHC}
\author{P.M. Ferreira~\footnote{ferreira@cii.fc.ul.pt},
R. Santos~\footnote{rsantos@cii.fc.ul.pt}}
\affiliation{Centro de F\'{\i}sica Te\'orica e Computacional, Faculdade de Ci\^encias,\\
Universidade de Lisboa, Avenida Professor Gama Pinto, 2, 1649-003
Lisboa, Portugal }
\date{April, 2006}
\maketitle \noindent {\bf Abstract.} We study the effects of a set
of dimension six flavor changing effective operators on several
processes of production of top quarks at the LHC. Namely,
$t\bar{t}$ production and associated production of a top and a
gauge or Higgs boson. Analytical expressions for the cross
sections of these processes are derived and presented.

\section{Introduction}

The large mass of the top quark~\cite{rev} makes it a natural
laboratory to search for new phenomena beyond those predicted by
the Standard Model (SM). One possible avenue of research consists
in using effective operators of dimensions larger than four to
parameterize the effects of any new physics. One advantage of
using this formalism is that one works in a model independent
manner. The complete set of dimension five and six operators that
preserve the gauge symmetries of the SM is quite large, and was
first obtained by Buchm\"uller and Wyler~\cite{buch}. This
methodology has been used by many authors to study the top quark.
For instance, in refs.~\cite{whis} contributions to top quark
physics arising from several types of dimension five and six
operators were studied. In ref.~\cite{saav} a detailed study of
the $W\,t\,b$ vertex was undertaken. Analysis of flavor changing
neutral currents in supersymmetric theories and models with two
Higgs doublets may be found in~\cite{fcnc}. For a recent study of
single top production in supersymmetric models see~\cite{sola} and
for a study on single top-quark production in flavor-changing Z'
models see~\cite{ari}. NLO and threshold corrections to top quark
flavor changing operators were obtained in~\cite{liu}. The four
fermion operator contributions to $t\bar{t}$ production were
studied in detail in~\cite{4f}.

Recently~\cite{nos1,nos2} we studied the effects on the
phenomenology of the top quark of a subset of operators of
dimension six - namely, those with contributions to strong flavor
changing neutral currents. The set chosen included several
operators studied by other authors, but which had not been
considered together before. We also benefitted greatly from
working in a fully gauge-invariant manner, taking advantage of the
equations of motion to eliminate several unknown effective
couplings. We considered both gluonic operators and four-fermion
ones. A detailed analysis of the contributions of these operators
in phenomena such as the top's width, its rare decays and the
cross section for single top production at the LHC was performed.
It was shown that the operator set we chose may have large
contributions to the single top production cross section at the
LHC, and that that channel is an excellent probe into the
existence of new physics.

In this paper we wish to analyze the effect of that set of
effective operators on other potentially interesting channels of
top production at the LHC, namely: top and anti-top production;
associated production of a top quark with a single gauge boson (a
photon, a $W$ or a $Z$ boson); and associated production of a top
quark and a Higgs boson. Our aim, as in
references~\cite{nos1,nos2}, is to produce analytical expressions
whenever possible, so that the results of this paper may be used
directly by our experimental colleagues in their Monte Carlo
simulations. This work is structured as follows: in
section~\ref{sec:eff} we will review the effective operator
formalism and the operators studied in refs.~\cite{nos1,nos2}.
Namely, we will explain the criteria behind that choice and the
role the equations of motion play in how many of them are truly
independent.  In section~\ref{sec:ttbar} we will study the effect
of our operator set in the production of $t\bar{t}$ pairs at the
LHC. In section~\ref{sec:tgauge} we will analyze the processes of
associated production of a top and a gauge boson and of a top and
a Higgs boson. In section~\ref{sec:num} we will present numerical
results for the cross sections of these processes at the LHC.
Finally, we will conclude in section~\ref{sec:conc} with an
overall analysis of the results.

\section{The effective operator approach}
\label{sec:eff}

A physical system rarely provides us enough information for a
complete description of its properties. A way to solve this
problem is to parameterize any physical effects not yet observed
by introducing an effective lagrangian with a set of new
interactions to be determined phenomenologically. This effective
lagrangian has the Standard Model as its low energy limit, and can
serve to represent the effect of any high-energy theory at a given
energy scale $\Lambda$. We write this lagrangian as 
\begin{equation}
{\cal L} \;\;=\;\; {\cal L}^{SM} \;+\; \frac{1}{\Lambda}\,{\cal L}^{(5)} \;+\;
\frac{1}{\Lambda^2}\,{\cal L}^{(6)} \;+\; O\,\left(\frac{1}{\Lambda^3}\right)
\;\;\; ,
\label{eq:l}
\end{equation}
where ${\cal L}^{SM}$ is the SM lagrangian and ${\cal L}^{(5)}$
and ${\cal L}^{(6)}$ are all of the dimension 5 and 6 operators
which, like ${\cal L}^{SM}$, are invariant under the gauge
symmetries of the SM. The ${\cal L}^{(5)}$ terms break baryon and
lepton number conservation, and usually are not considered. This
leaves us with the ${\cal L}^{(6)}$ operators. Some of these,
after spontaneous symmetry breaking, generate dimension five
terms. The complete list of effective operators was obtained
in~\cite{buch}. Our purpose, in this and previous
works~\cite{nos1, nos2} is to study flavor changing interactions,
restricted to the strong sector of the theory, involving a top
quark. Therefore, we choose operators with a single top quark,
that do not comprise gauge or Higgs bosons (except for those that
arise from covariant derivatives), and that involve some sort of
strong flavor changing interactions. Finally, we choose those
${\cal L}^{(6)}$ operators that have no sizeable impact on low
energy physics (by which we mean below the TeV scale).

Only two gluon operators survive these criteria which, in the
notation of ref.~\cite{buch}, are written as 
\begin{align}
{\cal O}_{uG} &= \;\;i\,\frac{\alpha_{ij}}{\Lambda^2}\,\left(\bar{u}^i_R\,
\lambda^a\, \gamma^\mu\,D^\nu\,u^j_R\right)\,G^a_{\mu\nu} \nonumber
\vspace{0.2cm} \\
{\cal O}_{uG\phi} &= \;\;\frac{\beta_{ij}}{\Lambda^2}\,\left(\bar{q}^i_L\,
\lambda^a\, \sigma^{\mu\nu}\,u^j_R\right)\,\tilde{\phi}\,G^a_{\mu\nu} \;\;\; .
\label{eq:op}
\end{align}
$q_L$ and $u_R$ are spinors (a left quark doublet and up-quark
right singlet of $SU(2)$, respectively), $\tilde{\phi}$ is the
charge conjugate of the Higgs doublet and $G^a_{\mu\nu}$ is the
gluon tensor. $\alpha_{ij}$ and $\beta_{ij}$ are complex
dimensionless couplings and the $(i,j)$ are flavor indices.
According to the criteria listed above, one of these indices must belong to the
third generation. After spontaneous symmetry breaking the neutral
component of the field $\phi$ acquires a vev
($\phi_0\,\rightarrow\,\phi_0\,+\,v$, with $v\,=\, 246/\sqrt{2}$
GeV) and the second of these operators generates a dimension five
term. The lagrangian for new physics is then given by 
\begin{align}
{\cal L}\;\; =&\;\;\; \alpha_{tu}\,{\cal O}_{tu}\;+\; \alpha_{ut}\,{\cal O}_{ut}
\;+\; \beta_{tu}\,{\cal O}_{tu\phi}\;+\;\beta_{ut}\,{\cal O}_{ut\phi}\;+\;
\mbox{h.c.} \nonumber \vspace{0.2cm} \\
 =& \;\;\;\frac{i}{\Lambda^2}\,\left[\alpha_{tu}\,\left(\bar{t}_R\,\lambda^a\,
\gamma^\mu \,D^\nu\,u_R\right)\;+\;  \alpha_{ut}\,\left(\bar{u}_R\,\lambda^a\,
\gamma^\mu\, D^\nu\,t_R\right)\right]\,G^a_{\mu\nu} \;\;\;+ \nonumber
\vspace{0.2cm} \\
 & \;\;\;\frac{v\,+\,h/\sqrt{2}}{\Lambda^2}\,\left[\beta_{tu}\,\left(\bar{t}_L\,\lambda^a
\, \sigma^{\mu\nu}\,u_R\right)\;+\; \beta_{ut}\,\left(\bar{u}_L\,\lambda^a\,
\sigma^{\mu\nu}\,t_R\right)\right]\,G^a_{\mu\nu} \;\;+\;\; \mbox{h.c.}
\;\;\;,
\label{eq:lf}
\end{align}
where $h$ is the SM Higgs boson.
This lagrangian describes new vertices such as $g\,\bar{t}\,u$,
 $g\,\gamma \, \bar{t}\,u$, $g\,Z\, \bar{t}\,u$ and $g\,h\,\bar{t}\,u$.
 There are, of course, analogous vertices involving the top quark,
instead of the anti-top one. We will also consider an analogous
lagrangian (with new couplings $\alpha_{tc}$, $\beta_{ct}$,
\ldots) for vertices of the form $g\,\bar{t}\,c$, $g\,\gamma \,
\bar{t}\,c$, $g\,Z\, \bar{t}\,c$ and $g\,h\,\bar{t}\,c$.  Notice
how the operators with $\beta$ couplings correspond to a
chromomagnetic momentum for the $t$ quark. Several extensions of
the SM, such as supersymmetry and two Higgs doublet models, may
generate contributions to this type of operator~\cite{chro}.

\begin{figure}[ht]
\epsfysize=6cm \centerline{\epsfbox{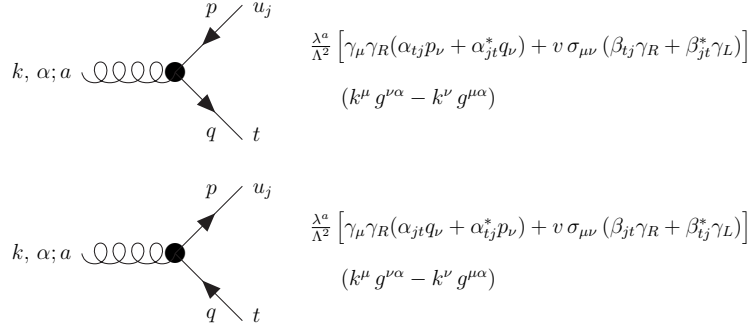}} \caption{Feynman
rules for anomalous gluon vertices.} \label{feynrul1}
\end{figure}

\begin{figure}[ht]
\epsfysize=4.5cm \centerline{\epsfbox{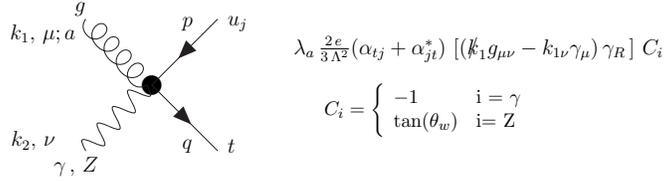}}
\caption{Feynman rules for anomalous gluon-$\gamma$ and gluon-Z
vertices.} \label{feynrul2}
\end{figure}

\begin{figure}[ht]
\epsfysize=4cm \centerline{\epsfbox{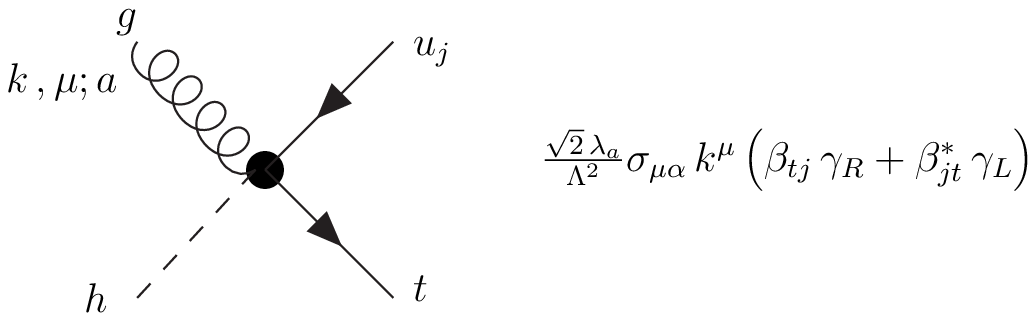}}
\caption{Feynman rules for anomalous gluon-h vertex.}
\label{feynrul3}
\end{figure}

The Feynman rules for these anomalous vertices are shown in
figs.~\eqref{feynrul1},~\eqref{feynrul2} and~\eqref{feynrul3}, with quark
momenta following the arrows and incoming gluon
momenta. The only operators that contribute to the Feynman rule in
fig.~\eqref{feynrul2}
are ${\cal O}_{ut}$ and ${\cal O}_{tu}$. They generate the vertices $g\,\gamma
\, \bar{t}\,u$ and $g\,Z\, \bar{t}\,u$ when we consider the electroweak
gauge fields present in the covariant derivatives of eq.~\eqref{eq:lf}.
On the other hand, the Feynman rule in fig.~\eqref{feynrul3} comes from the
operator
${\cal O}_{uG\phi}$, where the vev was replaced by the Higgs field. Of
course, we have considered analogous vertices involving the $c$ quark
instead of the $u$ one.

In ref.~\cite{nos1} we calculated the effect of these operators on
the width of the quark top. They allow for the decay
$t\,\rightarrow\,u\,g$ ($t\,\rightarrow \,c\,g$) (which is also
possible in the SM, but only at higher orders), and the
corresponding width is given by 
\begin{align}
\Gamma (t \rightarrow u g) &=\;  \frac{m^3_t}{12
\pi\Lambda^4}\,\Bigg\{ m^2_t \,\left|\alpha_{ut}  +
\alpha^*_{tu}\right|^2 \,+\, 16 \,v^2\, \left(\left| \beta_{tu}
\right|^2 + \left| \beta_{ut} \right|^2 \right) \;\;\; +
\vspace{0.3cm} \nonumber \\
 & \hspace{2.2cm}\, 8\, v\, m_t\,\mbox{Im}\left[ (\alpha_{ut}  + \alpha^*_{tu})
\, \beta_{tu} \right] \Bigg\} \label{eq:wid}
\end{align}
and an analogous expression for $\Gamma (t \rightarrow c g)$. In
this expression and in all results of this paper we have set all quark masses,
except that of the top, equal to zero. We performed the full calculations
and verified that the error introduced by this approximation is extremely small.
Notice how the top width~\eqref{eq:wid} depends on $\Lambda^{-4}$.
There are processes with a $\Lambda^{-2}$ dependence, namely the
interference terms between the anomalous operators and the SM
diagrams of single top quark production, via the exchange of a W
gauge boson - processes like
$u\,\bar{d}\,\rightarrow\,t\,\bar{d}$. They were studied in
ref.~\cite{nos2} in detail, and we discovered that, due to a
strong CKM suppression, the contributions from the anomalous
vertices are extremely small.

As was discussed in refs.~\cite{nos1,nos2}, the operators that compose the
lagrangian~\eqref{eq:lf} are not completely independent. If one performs
integrations by parts and uses the fermionic equations of
motion~\cite{buch,grz}, one obtains the following relations
between them: 
\begin{align}
{\cal O}^{\dagger}_{ut} &= {\cal O}_{tu}\;-\;\frac{i}{2} (\Gamma^{\dagger}_u\,
{\cal O}^{\dagger}_{u t \phi} \,+\, \Gamma_u \,{\cal O}_{t u \phi}) \nonumber \\
{\cal O}^{\dagger}_{ut} &= {\cal O}_{tu}\;-\;i\, g_s\, \bar{t}\, \gamma_{\mu}\,
\gamma_R\, \lambda^a\,u\, \sum_i  (\bar{u}^i\, \gamma^{\mu}\, \gamma_R\,
\lambda_a u^i\,+\, \bar{d}^i\, \gamma^{\mu}\, \gamma_R\, \lambda_a\, d^i)
\;\;\; ,
\label{eq:rel}
\end{align}
where $\Gamma_u$ are the Yukawa couplings of the up quark and
$g_s$ the strong coupling constant. In the second equation
four-fermion terms appear, which means that they have to be taken
into account in these studies. Indeed, their role was of great
importance for the processes studied in ref.~\cite{nos2}. In the
current paper, however, they will have no bearing in our results.
The most interesting thing about eqs.~\eqref{eq:rel}, which is in
fact a direct consequence of working in a fully gauge invariant
manner, is that they  tell us that there are two relations between
the several operators. This means that we are allowed to set two
of the couplings to zero. We have used this in ref.~\cite{nos2} to
simplify immensely the expressions obtained, by setting one of the
four-fermion couplings to zero, as well as making $\beta_{tu} =
\beta_{tc} = 0$. For consistency, we will make the same choice in
the current work, to allow for a direct comparison with the
results of~\cite{nos2}.

In refs.~\cite{nos1,nos2} we considered the contributions from the
anomalous flavor changing operators to single top production. In
particular, we calculated all processes with a top quark in the
final state, with a jet or isolated, stemming from either direct
top production or associated production with a gluon or a light
quark. We determined that the single top channel is an excellent
one for detection of new physics, as our calculations demonstrated
that one could obtain a significant increase in the cross section
of single top quark via the anomalous couplings relative to the SM
predicted values. We now wish to study the impact that these same
operators may have on other channels of top production, namely
$t\,\bar{t}$ production and the associated production of a top and
a gauge or Higgs boson. These are all channels of great physical
interest. In the former case, the LHC is expected to be a
veritable top-anti-top factory, with an estimated production of
around eight million top quark pairs per year and per experiment.
With such high statistics, a deviation from the SM prediction has,
{\em a priori}, a good chance of being detected. In the latter
case, the final state presents a very clear signal for
experiments. As such it should be easy to isolate it from the LHC
backgrounds.

At this point we must emphasize one important aspect: we are {\em
not} considering the most general set of operators for associated
top plus gauge or Higgs boson production processes. As mentioned
before, there are contributions from the electroweak dimension six
operators that we could have considered as well. However, that is
not our goal: we have established that the operator set we have
chosen, which corresponds to a specific type of possible new
physics (strong flavor changing interactions), may have a sizeable
impact on the single top channel. We now wish to verify if the
same operators might have important contributions to other
interesting channels. This is an important verification, to ensure
the consistency of the operator set chosen. If it predicts
significant increases on several physical processes but only some
of those are observed, then we will have a powerful clue that the
operators we chose do not tell the whole story of the new top
physics at the LHC. If, however, the observations are according to
the predictions arising from these operators, that will constitute
good evidence that they parameterize well whatever new physics
lies beyond the SM.

\section{Cross sections for $g\, g \,\rightarrow\,t\,\bar{t}$ and $q\,\bar{q}\,
\rightarrow\,t\,\bar{t}$}
\label{sec:ttbar}

\begin{figure}[ht]
\epsfysize=4.75cm \centerline{\epsfbox{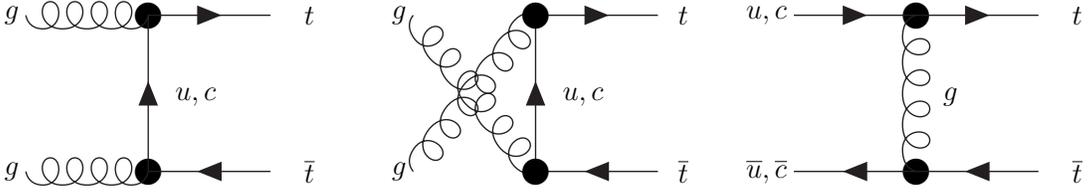}}
\caption{Feynman diagrams for $t \, \bar{t}$ production.}
\label{fig:gg}
\end{figure}

There are three Feynman diagrams contributing to the partonic
cross section of $t\,\bar{t}$ production, as is shown in
fig.~\eqref{fig:gg}. All of these diagrams contribute to the
process $p\, p \,\rightarrow\,t\,\bar{t}$ and interfere with the
(analogous) SM tree-level diagrams for the same processes. Notice
that, since each of these diagrams includes {\em two} anomalous
vertices, they will generate amplitudes of the order
$1/\Lambda^4$.
With the Feynman rules shown in fig.~\eqref{feynrul1} and the
corresponding SM Feynman rules, it is easy to obtain the
interference cross section for $g\, g \,\rightarrow\,t\,\bar{t}$,
given by
\begin{equation}
\frac{d\,\sigma(g\,g\rightarrow t\,\bar{t})}{dt}\;\; =\;\; -\,
\frac{g_s^2}{1536\,\pi \,\Lambda^4}\; \frac{F^1_{gg} \,
|\alpha_{ut}+\alpha_{tu}|^2 \, + \,F^2_{gg} \,
(|\beta_{ut}|^2+|\beta_{tu}|^2) \, + F^3_{gg} \, Im[\alpha_{ut} \,
\beta_{tu} -\alpha_{tu} \, \beta_{tu}^*] \, }{
    s^3\,\left( m_t^2 - t \right) \,t\,\left( m_t^2 - u \right) \,u} \;\;\; , \label{eq:gg}
\end{equation}
where 
\begin{align}
F^1_{gg}\,&=\,7\,m_t^{12}\,t - 23\,m_t^{10}\,t^2 +
        16\,m_t^8\,t^3 + 7\,m_t^{12}\,u -
        20\,m_t^{10}\,t\,u + 51\,m_t^8\,t^2\,u -73\,m_t^6\,t^3\,u+
\nonumber \vspace{0.3cm} \\
 &     \;\;\;\; 37\,m_t^4\,t^4\,u- 23\,m_t^{10}\,u^2 + 51\,m_t^8\,t\,u^2 -
        32\,m_t^6\,t^2\,u^2 - 31\,m_t^4\,t^3\,u^2 +
        35\,m_t^2\,t^4\,u^2 +
 \nonumber \vspace{0.3cm} \\
&      \;\;\;\; 2\,t^5\,u^2 +16\,m_t^8\,u^3-73\,m_t^6\,t\,u^3 -
31\,m_t^4\,t^2\,u^3 +
        42\,m_t^2\,t^3\,u^3 - 16\,t^4\,u^3 + 37\,m_t^4\,t\,u^4 +
        \nonumber \vspace{0.3cm} \\
& \;\;\;\; 35\,m_t^2\,t^2\,u^4 - 16\,t^3\,u^4 + 2\,t^2\,u^5 \,
 \nonumber \vspace{0.8cm} \\
F^2_{gg} \, &=\,\, 16 \, v^2 \, t \, u
             \left( 7\,m_t^6\,t - 15\,m_t^4\,t^2
             + 8\,m_t^2\,t^3 + 7\,m_t^6\,u - 26\,m_t^4\,t\,u +20\,m_t^2\,t^2\,u +
             \right.
             \nonumber \vspace{0.5cm} \\
 & \;\;\;\;
             \left.  t^3\,u - 15\,m_t^4\,u^2 +
             20\,m_t^2\,t\,u^2 - 16\,t^2\,u^2 + 8\,m_t^2\,u^3 + t\,u^3 \right) \,   \;\;\;
              \nonumber \vspace{0.8cm} \\
F^3_{gg} \, &=\,\, - \, 2 \, v \, m_t
             \left( 7\,m_t^{10}\,t - 23\,m_t^8\,t^2 +
      16\,m_t^6\,t^3 + 7\,m_t^{10}\,u - 52\,m_t^8\,t\,u +
      145\,m_t^6\,t^2\,u-
             \right.
             \nonumber \vspace{0.5cm} \\
 & \;\;\;\;
158\,m_t^4\,t^3\,u +
      60\,m_t^2\,t^4\,u - 23\,m_t^8\,u^2 +145\,m_t^6\,t\,u^2 - 258\,m_t^4\,t^2\,u^2 +
      136\,m_t^2\,t^3\,u^2 +
 \nonumber \vspace{0.3cm} \\
 & \;\;\;\;
             \left.   4\,t^4\,u^2 + 16\,m_t^6\,u^3 -
      158\,m_t^4\,t\,u^3 + 136\,m_t^2\,t^2\,u^3 - 64\,t^3\,u^3 +
      60\,m_t^2\,t\,u^4 + 4\,t^2\,u^4 \right) \;\;\;
. \nonumber \vspace{0.1cm} \\
\end{align}
For the $q\, \bar{q} \,\rightarrow\,t\,\bar{t}$ partonic channel we obtain
\begin{align}
 \vspace{0.1cm} \nonumber \\
\frac{d\,\sigma(q\,q\rightarrow t\,\bar{t})}{dt} \, & =\; -\,
\frac{g_s^2}{108\,\pi \,\Lambda^4\, s^3}\; \left(F^1_{qq} \, |\alpha_{tu}|^2
\, + \,F^2_{qq} \, |\alpha_{ut}| ^2 \, + \, F^3_{qq} \,
Re[\alpha_{ut} \, \alpha_{tu}]\, \right.
\nonumber \vspace{2.5cm} \\
 &     \;\;\;\; \left.
F^4_{qq}  \, (|\beta_{ut}|^2+|\beta_{tu}|^2) \, + \, F^5_{qq} \,
Im[\alpha_{ut} \, \beta_{tu}] \,  +  \, F^6_{qq} \, Im[\alpha_{tu}
\, \beta_{tu}^*] \right) \;\;\; ,
\end{align}
where 
\begin{align}
F^1_{qq}\,&=\,(m_t^2-t) \, u^{2} \, ;
\nonumber \vspace{0.8cm} \\
F^2_{qq} \, &=\,\, - \, (m_t^6 + t \, m_t^4 - 4 \, u \, m_t^4 -
      2 \, t^2 \, m_t^2 + u^2 \, m_t^2 - 6 \, t \, u \, m_t^2 + t \,
      u^2)\, ;
\nonumber \vspace{0.8cm} \\
F^3_{qq} \, &=\,\, - \, (m_t^6 - t\, m_t^4 - 4\, u\, m_t^4 +
      2\, t\, u\, m_t^2 - 2\, t\, u^2) \, ;
             \nonumber \vspace{0.8cm} \\
F^4_{qq} \, &=\,\, 8 \, (3 u \, m_t^2 + s^2 - u^2) \,  v^2 \, ;
             \nonumber \vspace{0.8cm} \\
F^5_{qq} \, &=\,\, 2 \, m_t \, v\, (m_t^4 - 5\, t\, m_t^2 +
        4 \, u \, m_t^2 + 4 \, t^2 + 12\, t\, u) \, ;
             \nonumber \vspace{0.8cm} \\
F^6_{qq} \, &=\,\, 2 \, m_t \, v\, (m_t^4 - t\, m_t^2 -6 \, u\,
m_t^2 +\, 4 \, t\, u) \, .
\end{align}
Despite the rather long expressions, notice that the dependence on
the $\{ \alpha\,,\,\beta\}$ anomalous couplings is quite simple.
We have kept all couplings in these expressions, but recall that,
due to the freedom allowed by the equations of motion, we are
allowed to set $\beta_{tu} = 0$. Finally, there are identical
expressions for the partonic cross sections involving the charm
anomalous couplings.

\section{Cross sections for associated $t \, \gamma$, $t \, Z$, $t \, W$ and
$t \, h$ production}
\label{sec:tgauge}

For the processes $q \, g \,\rightarrow\,t\,\gamma, \, Z$ there
are once more contributions from three diagrams. This process does
not occur in the SM at tree level. This is, thus, an order
$1/\Lambda^4$ process. In our previous papers, the top quark was
produced alongside with a quark or a gluon, and therefore detected
through a final state of t+jets.  Here, we have the production of
a top quark along with a gauge or Higgs boson, hopefully a much
``cleaner" signal for experimental detection. However, we must
recall that the final states of these channels will also include
jets, stemming either from initial and final state gluonic
radiation, or from remnants of the proton-proton collisions.
\begin{figure}[ht]
\epsfysize=4.5cm \centerline{\epsfbox{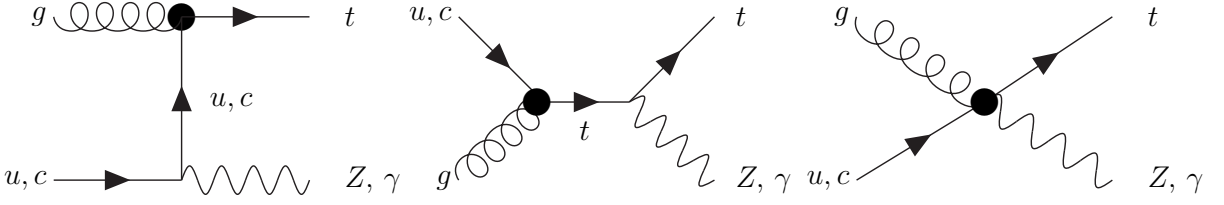}}
\caption{Feynman diagrams for the processes of $t \,\gamma$ and $t\,Z$
production}
\label{fig:gq}
\end{figure}
Using the Feynman rules from figs.~\eqref{feynrul1}
and~\eqref{feynrul2} we may obtain the cross section for
associated top plus photon production, given by
\begin{equation}
\frac{d\,\sigma(g\,q\rightarrow t\,\gamma)}{dt}\;\; =\;\;
\frac{e^2}{18 \, m_t^2 \, s^2 \, t \, (t+u)^2}\; (m_t^6-t \,
m_t^4+ s^2 \, m_t^2+3 \, s \, t \, m_t^2- 2 \, s^2 \, t) \, u \,
\, \Gamma (t \rightarrow q \, g) \;\;\; . \label{eq:gamma}
\end{equation}
where $\Gamma (t \rightarrow q \, g)$ stands for the decay width
of a top quark into a light up quark and a gluon. This result is
remarkably simple, and quite elegant. Similar expressions had been
obtained in refs.~\cite{nos1,nos2} for single top production in
the direct, gluon-gluon and gluon-quark channels. In fact, we
verified that every time there was a gluon in the initial or final
states the differential cross section was always proportional to a
partial decay width of the top. It is interesting to see the same
thing happening when a gluon is replaced by a photon.
Eq.~\eqref{eq:gamma} then establishes a very powerful link between
the rare decays of the top quark and the cross section for
associated top plus photon production.

Let us now consider the associated production of a top quark and a
$Z^0$ gauge boson. The calculation is similar to the top plus
photon channel, modulus the obvious kinematic differences, and the
different Feynman rules. We obtain, for the differential cross
section, the expression
\begin{equation}
\frac{d\,\sigma(g\,u\rightarrow t\,Z)}{dt}\;\; =\;\;  \frac{e^2 \,
m_t^2}{1728\,\pi \,\Lambda^4\, S^2_{2w}}\; \frac{F^1_{tZ} \,
|\alpha_{ut}+\alpha^*_{tu}|^2 \, + \,F^2_{tZ} \,
Im[(\alpha_{ut}+\alpha^*_{tu}) \, \beta_{tu} ]\, + \, F^3_{tZ} \,
|\beta_{tu}|^2 + F^4_{tZ} |\beta_{ut}|^2 \, \, }{m_z^2 \, s^2 \, t
\, (t+u)^2} \;\;\; , \label{eq:Z}
\end{equation}
where 

\begin{align}
F^1_{tZ}\,&=\,18\,m_t^2\,m_z^2\,s^2\,t +
  48\,m_t^2\,m_z^2\,s^2\,S_w^2\,t + 9\,s^2\,t^3 + 32\,m_t^6\,m_z^2\,S_w^4\,u +
\nonumber \vspace{0.3cm} \\
&      \;\;\;\;\;
  32\,m_t^2\,m_z^2\,s^2\,S_w^4\,u - 18\,m_z^2\,s^2\,t\,u+ 48\,m_z^2\,s^2\,S_w^2\,t\,u -
  \nonumber \vspace{0.3cm} \\
&   \;\;\;\;\; 32\,m_t^4\,m_z^2\,S_w^4\,t\,u +
  96\,m_t^2\,m_z^2\,s\,S_w^4\,t\,u -
  64\,m_z^2\,s^2\,S_w^4\,t\,u + 9\,s^2\,t^2\,u   \;\;\;\;
 \nonumber \vspace{0.8cm} \\
F^2_{tZ} \, &=\,\, \frac{4}{m_t} \left(18\,m_t^2\,m_z^2\,s^2\,t +
  48\,m_t^2\,m_z^2\,s^2\,S_w^2\,t + 9\,s^2\,t^3 +32\,m_t^6\,m_z^2\,S_w^4\,u +
   \right.
  \nonumber \vspace{0.5cm} \\
 & \;\;\;\;\;
  32\,m_t^2\,m_z^2\,s^2\,S_w^4\,u- 18\,m_z^2\,s^2\,t\,u + 48\,m_z^2\,s^2\,S_w^2\,t\,u -
             \nonumber \vspace{0.5cm} \\
 & \;\;\;\;
             \left.
  32\,m_t^4\,m_z^2\,S_w^4\,t\,u +
  96\,m_t^2\,m_z^2\,s\,S_w^4\,t\,u -
  64\,m_z^2\,s^2\,S_w^4\,t\,u + 9\,s^2\,t^2\,u \right) \;\;\;
              \nonumber \vspace{0.8cm} \\
F^3_{tZ} \, &=\,\,\frac{16}{m_t^2} \left( 54\,m_t^6\,m_z^2\,s -
54\,m_t^4\,m_z^2\,s^2 -
  48\,m_t^4\,m_z^2\,s^2\,S_w^2 +48\,m_t^2\,m_z^2\,s^3\,S_w^2 -
   \right.
              \nonumber \vspace{0.5cm} \\
 & \;\;\;\;
  54\,m_t^4\,m_z^2\,s\,t +54\,m_t^2\,m_z^2\,s^2\,t + 9\,s^2\,t^3 +18\,m_t^6\,m_z^2\,u - 54\,m_t^4\,m_z^2\,s\,u +
              \nonumber \vspace{0.5cm} \\
 & \;\;\;\;
  18\,m_t^2\,m_z^2\,s^2\,u -
  192\,m_t^4\,m_z^2\,s\,S_w^2\,u +192\,m_t^2\,m_z^2\,s^2\,S_w^2\,u -
  48\,m_z^2\,s^3\,S_w^2\,u +
            \nonumber \vspace{0.5cm} \\
 & \;\;\;\;
  32\,m_t^6\,m_z^2\,S_w^4\,u +
  32\,m_t^2\,m_z^2\,s^2\,S_w^4\,u -18\,m_t^4\,m_z^2\,t\,u +
  54\,m_t^2\,m_z^2\,s\,t\,u -
             \nonumber \vspace{0.5cm} \\
 & \;\;\;\;
 18\,m_z^2\,s^2\,t\,u -
  32\,m_t^4\,m_z^2\,S_w^4\,t\,u +
  96\,m_t^2\,m_z^2\,s\,S_w^4\,t\,u -64\,m_z^2\,s^2\,S_w^4\,t\,u +
             \nonumber \vspace{0.5cm} \\
 & \;\;\;
 \left.
  9\,s^2\,t^2\,u -48\,m_t^4\,m_z^2\,S_w^2\,u^2 +
  144\,m_t^2\,m_z^2\,s\,S_w^2\,u^2 -
  48\,m_z^2\,s^2\,S_w^2\,u^2
             \right) \;\;\;
              \nonumber \vspace{0.8cm} \\
F^4_{tZ} \, &=\,\, \frac{16}{m_t^2} \left(18\,m_t^2\,m_z^2\,s^2\,t
+
  48\,m_t^2\,m_z^2\,s^2\,S_w^2\,t + 9\,s^2\,t^3 +
             \right.
 \nonumber \vspace{0.5cm} \\
 & \;\;\;\;\; 32\,m_t^6\,m_z^2\,S_w^4\,u +
  32\,m_t^2\,m_z^2\,s^2\,S_w^4\,u -18\,m_z^2\,s^2\,t\,u + 48\,m_z^2\,s^2\,S_w^2\,t\,u -
 \nonumber \vspace{0.5cm} \\
 & \;\;\;\;            \left.
  32\,m_t^4\,m_z^2\,S_w^4\,t\,u +
  96\,m_t^2\,m_z^2\,s\,S_w^4\,t\,u -
  64\,m_z^2\,s^2\,S_w^4\,t\,u + 9\,s^2\,t^2\,u \right) \;\;\;
,
\end{align}
where we have used, for short, the convention $S_w = \sin(\theta_W)$ and
$S_{2w} = \sin(2\theta_W)$. The nice proportion to the top decay width is
destroyed in this expression, caused, no doubt, by the fact that the $Z^0$
boson, unlike the gluon or photon, is not massless.

We now turn to $t \, W$ production. Contrary to the $t \, Z$ and $t \,
\gamma$ processes, this one occurs at tree level in the SM. A discussion
about the problems related with $t \, W$ production and detection
can be found in~\cite{wt}.
\begin{figure}[ht]
\epsfysize=5cm \centerline{\epsfbox{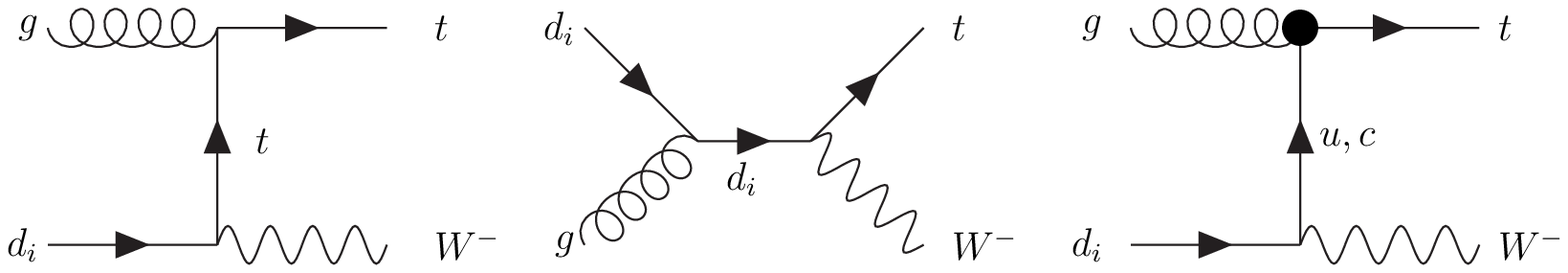}} \caption{Feynman
diagrams for $t \, W^-$ production.}
\label{fig:wt}
\end{figure}
In fig.~\eqref{fig:wt} we show the SM Feynman diagrams contributing
to this process, along with the single anomalous diagram that also
contributes to it. There is only one diagram because there are no contributions
of the type of fig.~\eqref{feynrul2}. The reason is simple: it is the covariant
derivative terms that give rise to the ``four-legged" diagram of
fig.~\eqref{feynrul2}, and that
derivative is acting on an $SU(2)$ gauge singlet.  Therefore, only the
hypercharge U(1) gauge field, $B_\mu$, contributes to the vertex. This means
that the operators ${\cal O}_{uG}$ do {\em not} give rise to any diagram of
the type of fig.~\eqref{feynrul2} for the $t \, W$ channels.

A simple calculation yields the SM tree level cross section, which reads
\begin{align}
\frac{d\,\sigma^{SM}(g\,d\rightarrow t\,W^-)}{dt}\; &
=\;\frac{e^2\,g_s^2 \, |V_{td}|^2  } {384\,
    m_w^2\,\pi \,s^3\,S_w^2\,( m_t^2 - t)^2} \, ( m_t^2\,u\,
       ( - t\,u   + m_t^2\,( 2\,t + u))  +
 \nonumber \vspace{1cm} \\
 & \;\;\;\;\;
      2\,m_w^2\,( 2\,m_t^4\,u +
         ( s + u ) \,( 2\,s^2 + 2\,s\,u +
            u\,( -2\,m_t^2 + u ))))  \, \, .
\label{eq:wsm}
\end{align}
The interference terms between the anomalous diagram and the SM
ones (for an internal q quark,  with $q\, =\, u\,,\,c$) were
computed in ref.~\cite{nos1}, and are given by
\begin{equation}
\frac{d\,\sigma^{INT}(g\,d\rightarrow t\,W^-)}{dt}\; =\;\frac{- \,
e^2\,g_s  \, |V_{td} \, V_{qd}|  \, m_t\,v
       \left(- m_t^2\,t\,u   +
        2\,m_w^2\,\left( s^2 + \left( m_t^2 + s \right) \,u \right)
        \right)
        }{48\,m_w^2\,\pi \,s^2\,S_w^2\,
    \left( m_t^2 - t \right) \,t} \,\, \frac{Re[\beta_{qt}]}{\Lambda^2} \,\,\, .
\label{eq:wint}
\end{equation}
Finally, the new anomalous term, which corresponds to the squared
amplitude of the anomalous diagram in fig.~\eqref{fig:wt}, is
given by
\begin{equation}
\frac{d\,\sigma^{NEW}(g\,d\rightarrow t\,W^-)}{dt}\; =\;\frac{e^2
\, |V_{qd}|^2 \left( m_t^2 - t \right) \,
    \left( s\,t + 2\,m_w^2\,u \right) \,v^2}{24\,m_w^2\,\pi
    \,s^2
    S_w^2\,t}  \,\, \frac{|\beta_{qt}|^2}{\Lambda^4} \, \, \, .
\end{equation}
Notice the dependence, in eqs.~\eqref{eq:wsm} and~\eqref{eq:wint}, on the
CKM matrix elements. We concluded, in~\cite{nos1}, that they were of vital
importance for the final results.

Finally, we consider the production of a top quark associated with
a SM Higgs boson. Like the $t \, Z$ and $t \, \gamma$ processes,
this process does not occur at tree level in the SM. The Feynman
diagrams are the same as for the top plus $\gamma$ or $Z$
processes. There are only two differences. First, we have to use
the four point Feynman rule in fig.~\eqref{feynrul3} instead of
the one in fig.~\eqref{feynrul2}. The ``four-legged" Feynman rule
involves now only the $\beta$ couplings instead of the $\alpha$
ones, which is natural, when you remember that the ${\cal O}_{uG}$
operators do not involve the Higgs boson in any way, whereas the
${\cal O}_{uG\phi}$ ones do. The second difference is the fact
that the diagram for top plus Higgs production analogous to the
second one in figure~\eqref{fig:gq} does {\em not} appear in these
calculations. The reason is very simple - the Higgs-quark vertex
is proportional to the mass of the quark in question and we have
taken, as explained earlier, $m_u\,=\,m_c\,=\,0$.

The cross section for $t \, h$ production can then be written as 
\begin{equation}
\frac{d\,\sigma(g\,u\rightarrow t\,h)}{dt}\;\; =\;\;
\left\{F^1_{th} \, |\alpha_{ut}+\alpha^*_{tu}|^2 \, + \,F^2_{th}
\, Im[(\alpha_{ut}+\alpha^*_{tu}) \, \beta_{tu} ]\, + \, F^3_{th}
\, (|\beta_{tu}|^2 + |\beta_{ut}|^2)\right\}\,\frac{1}{\Lambda^4}
\;\;\; , \label{eq:h}
\end{equation}
where
\begin{align}
F^1_{th}\,&=\frac{e^2\,m_t^2\,s^2\,
    \left( -\left( m_h^2\,m_t^2 \right)  - s\,t +
      m_t^2\,\left( 4\,s + t \right)  \right) }{48\,m_w^2\,
    {\left( m_t^2 - s \right) }^2\,S_w^2}   \;\;\;\;
 \nonumber \vspace{1.0cm} \\
F^2_{th} \, &=\,\frac{-\left( e\,m_t\,s^2\,
      \left( -\left( {\sqrt{2}}\,m_w\,\left( m_t^2 - s \right) \,
           S_w\,\left( 2\,m_t^2 - t \right)  \right)  +
        e\,m_t^2\,
         \left( -m_h^2 + 2\,\left( m_t^2 + s \right)  \right) \,v \right)
      \right) }{12\,m_w^2\,{\left( m_t^2 - s \right) }^2\,
    S_w^2} \;\;\;
              \nonumber \vspace{1.0cm} \\
F^3_{th} \, &=\,\, \frac{-1}{3\,m_w^2\,
    {\left( m_t^2 - s \right) }^2\,S_w^2} \left( s\,\left( -m_t^2 + s \right)
\,t\,
       \left( 2\,m_w^2\,\left( -m_t^2 + s \right) \,S_w^2 +
         2\,{\sqrt{2}}\,e\,m_t^2\,m_w\,S_w\,v -
         e^2\,m_t^2\,v^2 \right)  \right) +
\;
 \nonumber \vspace{1.0cm} \\
 &  \; \;\;\;\; m_t^2\,s\,\left( 2\,m_w^2\,{\left( m_t^2 - s \right)
}^2\, S_w^2 - 2\,{\sqrt{2}}\,e\,m_w\,
        \left( m_t^4 - s^2 \right) \,S_w\,v +
       e^2\,\left( -\left( m_h^2\,s \right)  +
          {\left( m_t^2 + s \right) }^2 \right) \,v^2 \right)  \; .
\end{align}

\section{Results for the integrated cross sections}
\label{sec:num}

\subsection{$t \, \bar{t}$ production}

The cross section for the gluon-gluon channel is identical for the
processes with anomalous couplings of the $u$ or $c$ quarks. For
the quark--antiquark cross section via a $c$ quark, the numerical
results are extremely small, and we do not present them. We have
used throughout this work the CTEQ6 parton density functions
(pdfs)~\cite{cteq6} and included a cut of 15 GeV on the transverse
momentum of the partons in the final state. This will allow a
direct comparison with the results of reference~\cite{nos2}, where
a similar cut was considered to help remove collinear and soft
singularities in the gluon-quark processes. Finally, for this
particular process, we have taken the factorization scale $\mu_F$
equal to twice the mass of the top quark. As was mentioned in
refs.~\cite{nos1,nos2}, this will produce smaller values of the
cross sections than we would obtain if we had, for instance, set
$\mu_F$ equal to the partonic center-of-mass energy. With these
specifications, the results we obtain are 
\begin{align}
\sigma_{p\,p\,\rightarrow\,g\,g\,\rightarrow\,t\bar{t}} &=\; \left\{-0.4 \,
|\alpha_{ut}+\alpha_{tu}|^2 \, + \, 7.6 \,
(|\beta_{ut}|^2+|\beta_{tu}|^2) \, + 9.1 \, Im[\alpha_{ut} \,
\beta_{tu} -\alpha_{tu} \, \beta_{tu}^*] \,
\right\}\,\frac{1}{\Lambda^4}
\;\mbox{pb} \vspace{0.3cm}\nonumber \\
\sigma_{p\,p\,\rightarrow\,q\,\bar{q}\,\rightarrow\,t \, \bar{t}} &=\;
\left\{\, -0.2 \, |\alpha_{tu}|^2 \, - \, 0.4 \, |\alpha_{ut}| ^2 \, + \, 0.5 \,
Re[\alpha_{ut} \, \alpha_{tu}]\, - \, 0.5 \,
(|\beta_{ut}|^2+|\beta_{tu}|^2) \, \, \right.
           \nonumber \vspace{0.5cm} \\
 &
\left. \;\;\;\;  \;\;  - \, 0.6 \, Im[\alpha_{ut} \, \beta_{tu}]
\, - \, 0.1 \, Im[\alpha_{tu} \, \beta_{tu}^*]
 \right\}\,
\frac{1}{\Lambda^4}\;\mbox{pb} \;\;\; .
\end{align}

So the total results for the proton-proton cross sections are
\begin{align}
 \sigma^{(u)}_{p\,p\,\rightarrow\,t \,
\bar{t}} &=\; \left\{\, -0.6 \, |\alpha_{tu}|^2 \, - \, 0.8 \,
|\alpha_{ut}| ^2 \, - \, 0.3 \, Re[\alpha_{ut} \, \alpha_{tu}]\, +
\, 7.1 \, (|\beta_{ut}|^2+|\beta_{tu}|^2) \, \, \right.
           \nonumber \vspace{0.5cm} \\
 &
\left. \;\;\;\;  \;\;  + \, 8.5 \, Im[\alpha_{ut} \, \beta_{tu}]
\, + \, 9.0 \, Im[\alpha_{tu} \, \beta_{tu}^*]
 \right\}\,
\frac{1}{\Lambda^4}
\;\mbox{pb} \vspace{0.3cm}\nonumber \\
\sigma^{(c)}_{p\,p\,\rightarrow\,t\bar{t}} &=\; \left\{-0.4 \,
|\alpha_{ct}+\alpha_{tc}|^2 \, + \, 7.6 \,
(|\beta_{ct}|^2+|\beta_{tc}|^2) \, + 9.1 \, Im[\alpha_{ct} \,
\beta_{tc} -\alpha_{tc} \, \beta_{tc}^*] \,
\right\}\,\frac{1}{\Lambda^4} \;\mbox{pb} \;\;\; .
\end{align}

\subsection{$t \, \gamma$ and $t \, Z$ production}

The results for top plus $\gamma$ production are particulary
simple, as the cross sections are proportional to the top decay
width to an up-type quark plus a gluon. The pdf suppression of the
$c$ quarks, however, makes the corresponding contributions to the
cross section extremely small, which is why we only present the
$u$ quark terms. For this channel, we chose $\mu_F \,=\, m_t$. The
proton-proton cross sections for top plus gamma production are
then given by
\begin{align}
\sigma^{(u)}_{p\,p\,\rightarrow\,t \, \gamma} \;\; &=\;\; 228 \,
\Gamma (t \rightarrow u \, g) \;|V_{tb}|^2 \;\mbox{pb}\; ;
\vspace{0.5cm}\nonumber \\
\sigma^{(u)}_{p\,p\,\rightarrow\,\bar{t} \, \gamma} \;\; &=\;\; 32.6
\, \Gamma (t \rightarrow u \, g) \; |V_{tb}|^2 \;\mbox{pb}\; .
\label{eq:gamma1}
\end{align}
We have also presented the cross section for anti-top plus gamma
production. To obtain this quantity we simply used the
differential cross section for the $t\,+\,\gamma$ channel,
eq.~\eqref{eq:gamma}, since there is no difference, in terms of
effective operators, between both processes. The different numbers
in eqs.~\eqref{eq:gamma1}, then, arise solely from different pdf
contributions (namely the $u$ and $\bar{u}$ quarks).

For the $t\,Z$ processes the expressions we obtain, after integrating on the
pdfs (with $\mu_F\,=\,m_t\,+\,m_Z$), are
\begin{align}
\sigma^{(u)}_{p\,p\,\rightarrow\,t \, Z} \;\; &=\;\; \left\{ \, 4.0
\, |\alpha_{ut}+\alpha^*_{tu}|^2 \, + \, 32.1 \,
Im[(\alpha_{ut}+\alpha^*_{tu}) \, \beta_{tu} ]\, + \, 63.8 \,
|\beta_{tu}|^2 + 65.3 |\beta_{ut}|^2 \right\} \,
\frac{1}{\Lambda^4} \; \mbox{pb}  ;
\vspace{0.5cm}\nonumber \\
\sigma^{(u)}_{p\,p\,\rightarrow\, \bar{t} \, Z} \;\; &=\;\; \left\{
\, 0.4 \, |\alpha_{ut}+\alpha^*_{tu}|^2 \, + \, 3.4 \,
Im[(\alpha_{ut}+\alpha^*_{tu}) \, \beta_{tu} ]\, + \, 6.7 \,
|\beta_{tu}|^2 + 7.0 |\beta_{ut}|^2 \right\} \,
\frac{1}{\Lambda^4} \; \mbox{pb} ;
\vspace{0.5cm}\nonumber \\
\sigma^{(c)}_{p\,p\,\rightarrow\,t \, Z} \;\; &=
\sigma^{(c)}_{p\,p\,\rightarrow\, \bar{t} \, Z}\;\; = \left\{ \,
0.2\, |\alpha_{ct}+\alpha^*_{tc}|^2 \, + \, 1.6 \,
Im[(\alpha_{ct}+\alpha^*_{tc}) \, \beta_{tc} ]\, + \, 3.2 \,
|\beta_{tc}|^2 + 3.4 |\beta_{ct}|^2 \right\} \,
\frac{1}{\Lambda^4} \; \mbox{pb}. \label{eq:Z1}
\end{align}

\subsection{$t \, W$ production}

With the SM diagrams of figure~\eqref{fig:wt}, we obtain the
expected tree-level result for $\sigma_{p \, p \rightarrow
t\,W^-}$, which is about 30 pb. Unlike the previous cases of
production of a top quark alongside with a gauge boson, we now
have interference terms between the SM diagrams and the anomalous
ones, which are of order $\Lambda^{-2}$. We obtained these
interference cross sections in ref.~\cite{nos1} and present them
here again. For the $u$ quark anomalous couplings we have
($\mu_F\,=\,m_t\,+\,m_W$)
\begin{align}
\sigma^{INT}(p\,p \rightarrow t\,W^-)\; &=\; 0.031 \;
Re[\beta_{ut}]\; \frac{1}{\Lambda^2} \;\; \mbox{pb} ;
\vspace{0.5cm}\nonumber \\
\sigma^{INT}(p\,p \rightarrow \bar{t} \,W^+)\; &=\; 0.022 \;
Re[\beta_{ut}]\; \frac{1}{\Lambda^2} \; \;\mbox{pb} ,
\end{align}
whereas for the $c$ quark couplings the interference cross sections are
\begin{align}
\sigma^{INT}(p\,p \rightarrow t\,W^-)\; &=\; 0.065\;
Re[\beta_{ct}] \; \frac{1}{\Lambda^2} \; \; \mbox{pb} ;
\vspace{0.5cm}\nonumber \\
\sigma^{INT}(p\,p \rightarrow \bar{t} \,W^+)\; &=\; 0.063 \;
Re[\beta_{ct}]  \; \frac{1}{\Lambda^2} \; \mbox{pb} .
\end{align}
These cross sections have extremely small coefficients, which is
due to a double CKM cancellation, as we have discussed in
ref.~\cite{nos1}. This cancellation makes the contribution to the
cross section arising from the square of the anomalous diagram the
largest, as long as the scale of new physics isn't too large
(namely, the interference terms for the u coupling are negligible
as long as $\Lambda/\sqrt{|\beta_{ut}|} < 45$ GeV).

The new contributions to $t\,W$ production are then given by, for the $u$
couplings,
\begin{align}
\sigma^{NEW}(p\,p \rightarrow t\,W^-)\; &=\; 63.4 \,
|\beta_{ut}|^2 \frac{1}{\Lambda^4} \; \mbox{pb}
\vspace{0.5cm}\nonumber \\
\sigma^{NEW}(p\,p \rightarrow \bar{t} \,W^+)\; &=\; 18.0 \,
|\beta_{ut}|^2 \frac{1}{\Lambda^4} \; \mbox{pb}\;\;\;,
\end{align}
and, for the $c$ quark couplings,
\begin{align}
\sigma^{NEW}(p\,p \rightarrow t\,W^-)\; &=\; 14.2 \,
|\beta_{ct}|^2 \frac{1}{\Lambda^4} \; \mbox{pb}
\vspace{0.5cm}\nonumber \\
\sigma^{NEW}(p\,p \rightarrow \bar{t} \,W^+)\; &=\; 11.9 \,
|\beta_{ct}|^2 \frac{1}{\Lambda^4} \; \mbox{pb}\;\;\; .
\end{align}

It is interesting to notice that because the SM process occurs
mostly through a $g \, b$ initial state, and the pdfs for a $b$
quark and a $\bar{b}$ quark are essentially identical, there is
almost no difference in $t$ and $\bar{t}$ production. That is,
$\sigma^{SM}(t\,W^-)\, - \, \sigma^{SM}(\bar{t} \,W^+)\; \approx\;
0$. However, the interference terms and the new ones receive
contributions from all quarks leading to a difference
\begin{align}
&\sigma^{INT}(t\,W^-)\, - \, \sigma^{INT}(\bar{t} \,W^+)\;=\; 0.09
Re[\beta_{ut}]  \; \frac{1}{\Lambda^2} \; \mbox{pb};
\vspace{0.5cm}\nonumber \\
&\sigma^{NEW}(t\,W^-)\, - \, \sigma^{NEW}(\bar{t} \,W^+)\;=\; 45.4
|\beta_{ut}|^2  \; \frac{1}{\Lambda^4} \; \mbox{pb}.
\end{align}
Therefore, this asymmetry could be a clear sign of new physics.
Moreover, it depends on only one of the anomalous couplings and, if no
asymmetry of this kind is observed, a stringent bound could be set on
$\beta_{ut}$.

\subsection{$t \, h$ production}

Finally, we consider the numerical results for associated top plus
Higgs boson production. The cross sections depend, of course, on
the unknown value of the Higgs mass. As we will observe, though,
that dependence is not a strong one. We will consider two values
for the Higgs mass, $m_h\,=\,120$ GeV (the preferred value from
the current experimental bounds, and a typical value of a
supersymmetric Higgs mass) and $m_h\,=\,300$ GeV. Once again, the
results we obtained for production of $t\,h$ via the $c$ quark are
too small, and we do not show them. Likewise, the pdf suppression
of the anti-up and anti-charm quark heavily suppress the
production of an anti-top quark and a Higgs boson. We are left
with $t\,h$ production via the $u$ quark, which, for $m_h\,=\,120$
GeV, reads
\begin{equation}
\frac{d\,\sigma(g\,u\rightarrow t\,h)}{dt}\;\; =\;\; \left\{5.9 \,
|\alpha_{ut}+\alpha^*_{tu}|^2 \, + \,23.6 \,
Im[(\alpha_{ut}+\alpha^*_{tu}) \, \beta_{tu} ]\, + \, 95.2 \,
(|\beta_{tu}|^2 + |\beta_{ut}|^2)\right\}\,\frac{1}{\Lambda^4} \;\;\; ,
\label{eq:h120}
\end{equation}
and, for $m_h\,=\,300$ GeV, we have
\begin{equation}
\frac{d\,\sigma(g\,u \rightarrow t\,h)}{dt}\;\; =\;\; \left\{ 3.2
\, |\alpha_{ut}+\alpha^*_{tu}|^2 \, + \,25.8 \,
Im[(\alpha_{ut}+\alpha^*_{tu}) \, \beta_{tu} ]\, + \, 52.1\,
(|\beta_{tu}|^2 + |\beta_{ut}|^2)\, \right\}\frac{1}{\Lambda^4}
\;\;\; . \label{eq:h300}
\end{equation}
We observe some variation of the coefficients of the $\alpha$ and
$\beta$ coefficients, the cross section for the larger Higgs mass
being, naturally, smaller. However, that variation is not a
dramatic one, which is perhaps due to the extremely high
center-of-mass energy available in the proton-proton collisions at
the LHC.

\section{Conclusions}
 \label{sec:conc}

In this section we perform a joint analysis of the results
obtained so far in this paper and those from our previous
works~\cite{nos1,nos2}. We have calculated all possible one and
two body decays, at the partonic level, originating from a set of
strong flavor changing operators satisfying our predefined
criteria. We now wish to observe if there is any correlation
between the cross sections of top quark production we computed. In
what follows we have used the liberty afforded to us by the
equations of motion to set $\beta_{tu}\,=\,\beta_{tc}\, =\,0$, to
simplify our calculations.

\begin{figure}[ht]
\epsfysize=9cm \centerline{\epsfbox{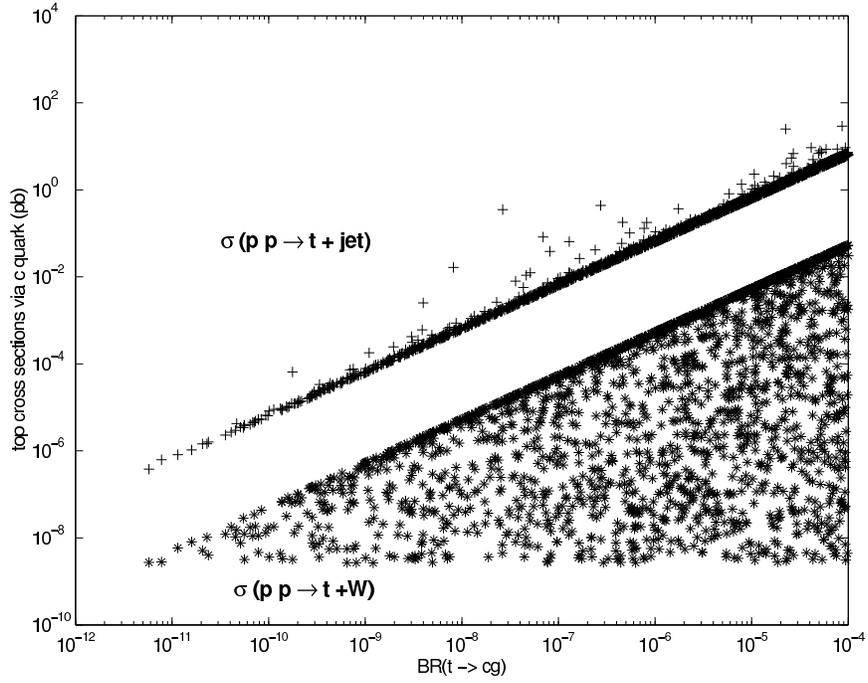}} \caption{$p \, p
\rightarrow t \, + \,  jet$ (+) and $p \, p \rightarrow t \, + \,
W$ (*) via a $c$ quark as a function of the branching ratio $BR(t
\rightarrow g \, c)$ .} \label{fig:sig_t_c}
\end{figure}

To investigate the dependence of the cross sections on the values
of the anomalous couplings, we generated random values for these,
and plotted the cross sections against the branching ratios of the
top quark for the decays $t \rightarrow g\,u$ and $t \rightarrow
g\,c$. Our rationale for doing this is a simple one: as was
discussed in refs.~\cite{chro,juan}, the top quark branching
ratios for these decays may vary by as much as eight orders of
magnitude, from $\sim\,10^{-12}$ in the SM to $\sim\,10^{-4}$ for
some supersymmetric models. This quantity, then, is a good measure
of whether any physics beyond that of the standard model exists.
\begin{figure}[ht]
\epsfysize=9cm \centerline{\epsfbox{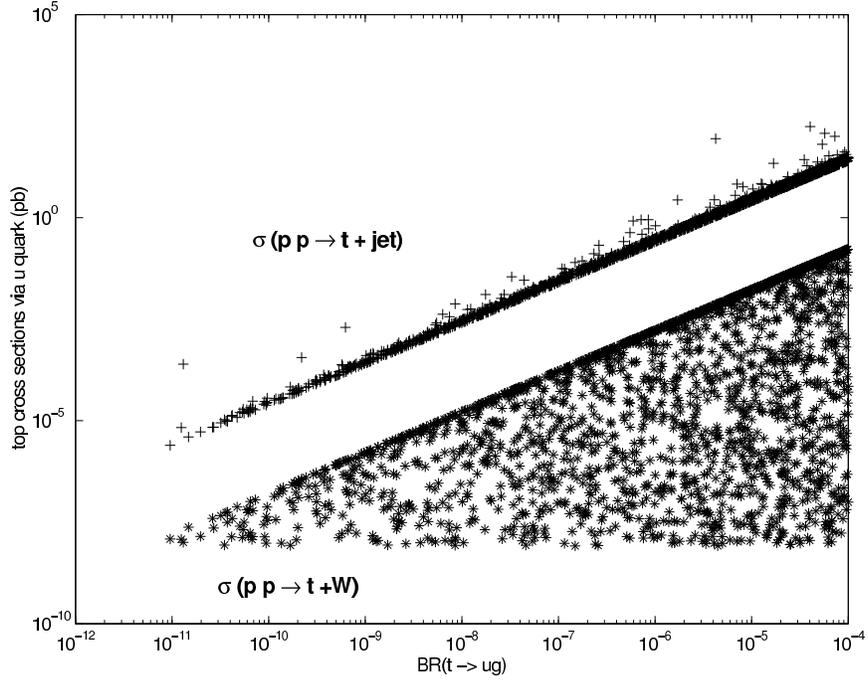}} \caption{Cross
sections for the processes $p\,p \rightarrow t \,+\, jet$
(crosses) and $p\,p \rightarrow t \, + \, W$ (stars) via an $u$
quark, as a function of the branching ratio $BR(t \rightarrow g \,
u)$. } \label{fig:sig_t_u}
\end{figure}
In fig.~\ref{fig:sig_t_u} we show the plot of the cross sections
for the processes $p \, p \rightarrow t\, + \,  jet$ and $p \,p
\rightarrow t\, +\, W$ via a $u$ quark versus the branching ratio
$BR(t \rightarrow g \, u)$. This plot was obtained by varying the
constants $\alpha$ and $\beta$ in a random way. Each combination
of $\alpha$ and $\beta$ originates a given branching ratio and a
particular value for each cross section. Obviously, another set of
points may generate the same value for the branching ratio but a
different value for the cross section, which justifies the
distribution of values of $\sigma(p\,p \rightarrow t\,+\,jet)$ and
$\sigma(p\,p \rightarrow t\,+\,W)$. We chose values of $\alpha$
and $\beta$ for which the branching ratio varies between its SM
value and the maximum theoretical supersymmetric value it may
assume. In fig.~\eqref{fig:sig_t_c} we show a similar plot, but
for top plus jet and top plus a $W$ boson production via a $c$
quark.

It is obvious from both fig.~\ref{fig:sig_t_u} and
fig.~\ref{fig:sig_t_c} that if the branching ratio is close to its
SM value there is no chance to observe new strong flavor changing
physics at the LHC. However, as we approach the larger values of
$10^{-4}$, the cross section for single top becomes visible and
the $W \, t$ cross section approaches 0.1 pb. Notice that the
$t\,W$ cross section is proportional to only one of the couplings,
which makes it  a very attractive observable - it may allow us to
impose constraints on a single anomalous coupling.

It should be noted that single top production depends also on the
contributions of the four fermion operators. Hence, even if the
branching ratios  $BR(t \rightarrow g \, u (c))$ are very small,
there is still the possibility of having a large single top cross
section with origin in the four fermion couplings. In figs.~\ref{fig:sig_t_u}
and~\ref{fig:sig_t_c} we did not consider this possibility, setting the
four-fermion couplings to zero. For a discussion on the four-fermion
couplings do see~\cite{nos2}.

\begin{figure}[ht]
\epsfysize=9cm \centerline{\epsfbox{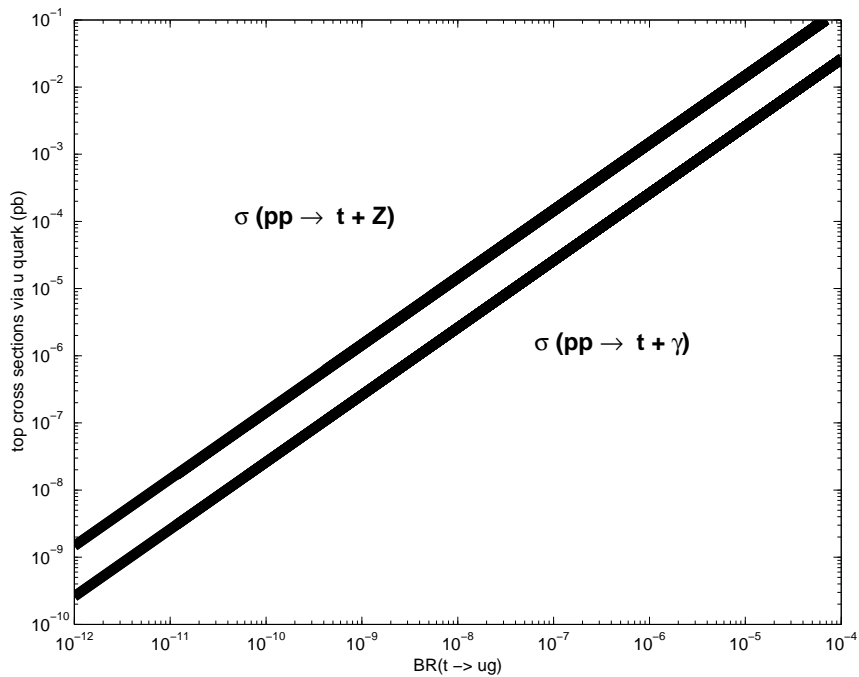}} \caption{Cross
sections for the processes $p \, p \rightarrow t \, + \,  Z$
(upper line) and $p \, p \rightarrow t \, + \,\gamma$ (lower line)
via a $u$ quark, as a function of the branching ratio $BR(t
\rightarrow g \, u)$ .} \label{fig:Zgamma_u}
\end{figure}

In fig.~\ref{fig:Zgamma_u} we plot the cross sections for $p \, p \rightarrow
t \, + \, Z$ and $p \, p \rightarrow t \, + \,\gamma$ via a $u$ quark, versus
the branching ratio $BR(t \rightarrow g \, u)$. The
equivalent plot with an internal $c$ quark is similar, but the
values for the cross section are much smaller. In this plot we can see
that both cross sections are very small in the range of $\{\alpha\,\beta\}$
considered. These results imply that their contribution will hardly be seen at
the LHC, unless the values for the branching ratio are peculiarly large.

\begin{figure}[ht]
\epsfysize=9cm \centerline{\epsfbox{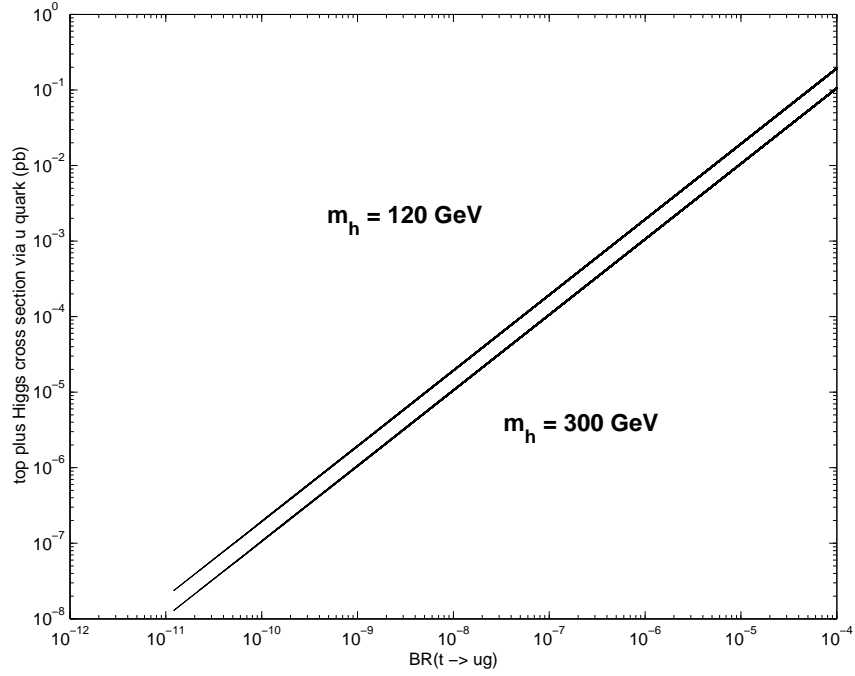}} \caption{Cross
section of the process $p \, p \rightarrow t \, + \,  h$ via a $u$
quark versus the branching ratio $BR(t \rightarrow g \, u)$ for
$m_h\,=\,120$ GeV and $m_h\,=\,300$ GeV.} \label{fig:top_higgs}
\end{figure}

The same, in fact, could be said for $p \, p \rightarrow t \, + \,
h$. In fig.~\ref{fig:top_higgs} we present a plot for this cross
section, again as a function of the branching ratio of $t
\rightarrow g \, u$, for two values of the Higgs mass. We readily
see that, even for the smallest allowed SM Higgs mass, the values
are very small. The same holds true for the process involving the
anomalous couplings of the $c$ quark.

The smallness of the effects of these operators in the several cross sections
holds true, as well, for the top--anti-top channel. In this case, even for
a branching ratio $BR(t \rightarrow g \, u) \,\simeq\,10^{-4}$, the
contributions to the cross section $\sigma(p \, p \rightarrow t \,\bar{t})$
do not exceed, in absolute value, one picobarn. They may be positive or
negative, but always extremely small.

In conclusion, the effective operators we have considered in this
paper and references~\cite{nos1,nos2} are extremely constrained in
their impact on the several channels of top quark production.
Namely, figs.~\eqref{fig:sig_t_u} through~\eqref{fig:top_higgs}
illustrate that, with the exception of the cross section for
production of a single top plus a jet, the other channels are
expected to have anomalous contributions which are probably too
small to be observed at the LHC. Thus, if there are indeed strong
flavor changing neutral current effects on the decays of the top
quark, the results of this paper show that their impact will be
restricted to a single channel, single top plus jet production. It
is entirely possible, according to our results, to have an excess
in the cross section $\sigma(p \, p \rightarrow t \, +
\,\mbox{jet})$ arising from new physics described by the operators
we have considered here, at the same time obtaining results for
the production of a top quark alongside a gauge and Higgs boson,
or for $t\bar{t}$ production, which are entirely in agreement with
the SM predictions. This reinforces the conclusion of
reference~\cite{nos2}: that the cross section for single top plus
jet production is an excellent probe for the existence of new
physics beyond that of the SM. It is a channel extremely sensitive
to the presence of that new physics, and boasts a significant
excess in its cross section, whereas many other channels involving
the top quark remain unchanged. Nevertheless, we are encouraged by
the fact that it may still be possible to use some of these
unchanged channels, such as top plus $W$ production, to constrain
the $\beta$ parameters, through the study of asymmetries such as
$\sigma(p \, p \rightarrow t\,W^-)\, - \, \sigma(p \, p
\rightarrow \bar{t} \,W^+)$.

\vspace{0.25cm} {\bf Acknowledgments:} Our thanks to our
colleagues from LIP for valuable discussions. Our further thanks
to Ant\'onio Onofre for a careful reading of the manuscript. This
work is supported by Funda\c{c}\~ao para a Ci\^encia e Tecnologia
under contract POCI/FIS/59741/2004. P.M.F. is supported by FCT
under contract SFRH/BPD/5575/2001.


\newpage




\begin{thebibliography}{99}
\bibitem{rev} M. Beneke {\em et al}, {\bf hep-ph/0003033};

D. Chakraborty, J. Konigsberg and D. Rainwater, {\em Ann. Rev.
Nucl. Part. Sci.} {\bf 53} (2003) 301;

W. Wagner, {\em Rept. Prog. Phys.} {\bf 68} (2005) 2409.





\bibitem{buch} W. Buchm\"uller and D. Wyler, {\em Nucl. Phys.} {\bf B268} (1986)
621.
\bibitem{whis} E. Malkawi and T. Tait, {\em Phys. Rev.} {\bf D54} (1996) 5758;

T. Han, K. Whisnant, B.L. Young and X. Zhang, {\em Phys. Lett.}
{\bf B385} (1996) 311;

T. Han, M. Hosch, K. Whisnant, B.L. Young and X. Zhang, {\em Phys.
Rev.} {\bf D55} (1997) 7241;

K. Whisnant, J.M. Yang, B.L. Young and X. Zhang, {\em Phys. Rev.}
{\bf D56} (1997) 467;

M. Hosch, K. Whisnant and B.L. Young, {\em Phys. Rev.} {\bf D56}
(1997) 5725;

T. Han, M. Hosch, K. Whisnant, B.L. Young and X. Zhang, {\em Phys.
Rev.} {\bf D58} (1998) 073008;

K. Hikasa, K. Whisnant, J.M. Yang and B.L. Young, {\em Phys. Rev.}
{\bf D58} (1998) 114003.





\bibitem{saav}F. del \`Aguila and J.A. Aguilar-Saavedra, {\em Phys. Rev.} {\bf
D67} (2003) 014009.




\bibitem{fcnc}
T. Tait and C. P. Yuan, {\em Phys. Rev.} {\bf D63}, (2001) 014018;

D. O. Carlson, E. Malkawi, and C. P. Yuan, {\em Phys. Lett.} {\bf
B337}, (1994) 145;

G. L. Kane, G. A. Ladinsky, and C. P. Yuan, {\em Phys. Rev.} {\bf
D45}, (1992) 124;

T. G. Rizzo, {\em Phys. Rev.} {\bf D53}, (1996) 6218;

T. Tait and C. P. Yuan, {\em Phys. Rev.} {\bf D55}, (1997) 7300;

A. Datta and X. Zhang, {\em Phys. Rev.} {\bf D55}, (1997) 2530;

E. Boos, L. Dudko, and T. Ohl, {\em Eur. Phys. J.} {\bf C11},
(1999) 473;

D. Espriu and J. Manzano, {\em Phys. Rev.} {\bf D65}, (2002)
073005;

J.L. Diaz-Cruz, H.-J. He, C.P. Yuan, {\em Phys. Lett.} { \bf B530}
(2002) 179;

H.-J. He, C.P. Yuan, {\em Phys. Rev. Lett.} { \bf 83} ( 1999) 28.



\bibitem{sola} G. Eilam, M. Frank and I. Turan, {\bf hep-ph/0601253};

J. Guasch, W. Hollik, S. Pe\~naranda and J. Sol\`a, {\bf
hep-ph/0601218}.


\bibitem{ari}
  A.~Arhrib, K.~Cheung, C.~W.~Chiang and T.~C.~Yuan,
  arXiv:hep-ph/0602175.


\bibitem{liu} J. J. Liu, C. S. Li, L. L. Yang and L. G. Jin, {\em Phys. Rev.}
{\bf D72} (2005) 074018;

L.L. Yang, C. S. Li, Y. Gao and J. J. Liu, {\bf hep-ph/0601180}.
\bibitem{4f} C.T. Hill and S.J. Parke, {\em Phys. Rev.} {\bf D49} (1994) 4454;

G. J. Gounaris, D. T. Papadamou, F. M. Renard, {\em Z. Phys.} {\bf
C76} (1997) 3 33.


\bibitem{nos1} P.M. Ferreira and R. Santos,  {\em Phys. Rev.} {\bf D73} (2006)
054025.

\bibitem{nos2} P.M. Ferreira, O. Oliveira and R. Santos,  {\em Phys. Rev.}
{\bf D73} (2006) 034011;

\bibitem{chro} B. Grzadkowski, J.F. Gunion and P. Krawczyk, {\em Phys. Lett.}
{\bf B268} (1991) 106;

G. Eilam, J.L. Hewett and A. Soni, {\em Phys. Rev.} {\bf D44}
(1991) 1473;

T.P. Cheng and M. Sher, {\em Phys. Rev.} {\bf D35} (1987) 3484;

L.J. Hall and S. Weinberg, {\em Phys. Rev.} {\bf D48} (1993) R979.


\bibitem{grz} B. Grzadkowski, Z. Hioki, K. Ohkuma and J. Wudka, {\em Nucl.
Phys.} {\bf B689} (2004) 108.

\bibitem{wt}
Z. Sullivan, {\em Phys. Rev.} {\bf D70} (2004) 114012.

\bibitem{cteq6} J. Pumplin {\em et al}, {\em JHEP} {\bf 0207} (2002) 012.


\bibitem{juan} M.E. Luke and M.J. Savage, {\em Phys. Lett.} {\bf B307} (1993)
387.





\end{thebibliography}
\end{document}